# Magnetic functionalization and Catalytic behavior of magnetic nanoparticles during laser photochemical graphitization of polyimide

Abhishek Sarkar[1], Ho-won Noh[2], Ikenna C. Nlebedim[1], Pranav Shrotriya[2]

1. Critical Materials Institute, Ames Laboratory, Ames, IA – 50011.
2. Department of Mechanical Engineering, Iowa State University, Ames, IA – 50011.

**Abstract:**

We report laser-assisted photochemical graphitization of polyimides (PI) into functional magnetic nanocomposites using laser irradiation of PI in the presence of magnetite nanoparticles (MNP). PI Kapton sheets covered with MNP were photochemically treated under ambient conditions using a picosecond pulsed laser (1064nm) to obtain an electrically conductive material. Scanning electron microscopy of the treated material revealed layered magnetic nanoparticles/graphite nanocomposite structure (MNP/graphite). Four probe conductivity measurements indicated that nanocomposite has an electrical conductivity of 1550±60 S/m. Superconducting quantum interference device (SQUID) magnetometer-based magnetic characterization of the treated material revealed an anisotropic ferromagnetic response in the MNP/graphite nanocomposite compared to the isotropic response of MNP. Raman spectroscopy of MNP/graphite nanocomposite revealed a four-fold improvement in graphitization, suppression in disorder, and decreased nitrogenous impurities compared to the graphitic material obtained from laser treatment of just PI sheets. X-ray photoelectron spectroscopy, x-ray diffraction, and energy-dispersive x-ray spectroscopy were used to delineate the phase transformations of MNP during the formation of MNP/graphite nanocomposite. Post-mortem characterization indicates a possible photocatalytic effect of MNP during MNP/graphite nanocomposite formation. Under laser irradiation, MNP transformed from the initial $Fe_3O_4$ phase to $\gamma\text{-}Fe_2O_3$ and $Fe_5C_2$ phases and acted as nucleation spots to catalyze the graphitization process of PI.

**Introduction**

Graphite and graphitic materials possess low shear resistance (soft), excellent thermal and electrical conductivity, high stiffness and strength, thermal stability at extreme temperatures (>3600°C), and selective chemical reactivity. These properties enable diverse applications of graphitic carbon[1], including electronics, mechanical components (lubricant, metallurgical, etc.), energy storage (batteries), and energy generation (nuclear)[2]. Natural graphite is the most common resource utilized in industry. However, the increasing demand and deficit of high-quality natural graphite, especially within the US[3], severely impacts the supply chain. Challenges with artificial graphite formation involve extreme temperature (>2000°C) thermal decomposition of polymers under inert atmospheric conditions. This results in 25% process efficiency and energy requirement of 4.5kWh/kg[4], exceeding the natural graphite price by up to 10X.

Laser-induced reduction of polymers into graphite provides a unique perspective towards room temperature and energy-efficient production of artificial graphite[5]. The concept of laser-induced transformation of diamond to graphite was established in the early 2000s[6]. In 2014, Lin et al.[7] reported the laser-induced transformation of polymeric films (polyimide, PI) into a 3D porous graphene structure (laser-induced graphene, LIG) using a $CO_2$ microsecond laser. Initial studies of LIG transformation were conducted on PI and polyether imide (PEI) that contain nitrogen groups and aromatic chains[5,7,8]. These polymers were selected due to their existing hexagonal ring structure and the presence of nitrogen within the polymerizing group. The ring structure plays a crucial role in the $sp^2$-hybridized stacking of the graphene layers, while the nitrogen creates a protective blanket to prevent incineration during the lasing process. Nevertheless, the current LIG graphitization process with a microsecond $CO_2$ pulsed laser induces a disordered hexagonal structure dominated by heptagonal and pentagonal irregularities[5] primarily due to a photothermal dominant conversion mechanism.

The structure and dynamics of the surface treated by pulsed-laser irradiation strongly depend on the laser pulse timescale[6,9,10]. Upon incidence, the laser pulse transfers its energy to the electron system. Typically, the electron-lattice relaxation time is in the order of picoseconds ($10^{-12}$ s)[11]. The electron transfers its kinetic energy into the lattice to bring the system to a thermal equilibrium (photothermal) for nanosecond and large pulse-width lasers. With ultrashort pulsed (pico and femtosecond) lasers, the laser energy is transferred to the electrons in a shorter time than the electron-lattice relaxation period. Thus, the energy remains within the electron gas (photochemical) and is not dissipated within the lattice. The electron system (hot) is expected to stay in equilibrium, while the lattice is cold. Compared to a nanosecond or microsecond laser, which can generate in the order of >2000 K in the lattice, a pico-/femtosecond can generate >10,000 K of equivalent temperature in the electron system[6].

LIGs have been functionalized with nanomaterials to enable applications in energy storage and sensors[5,8], micro supercapacitors[12,13], electrocatalysts[7,14,15], and other sensors[16,17]. However, to our knowledge, the synthesis of magnetic-functionalized graphitic materials using laser-assisted transformation of polymers has not been reported. In this work, we report the photochemical synthesis of magnetic nanoparticle/graphite nanocomposite using ultrashort pulse picosecond lasers and the photocatalytic behavior of magnetite ($Fe_3O_4$) nanoparticles on the conversion process (Fig.1). The laser-induced graphitization and magnetic behavior of the composite is analyzed. Post-mortem characterization of morphology, phase evolution, and graphitization are performed to elucidate and understand the possible conversion pathways during laser processing.

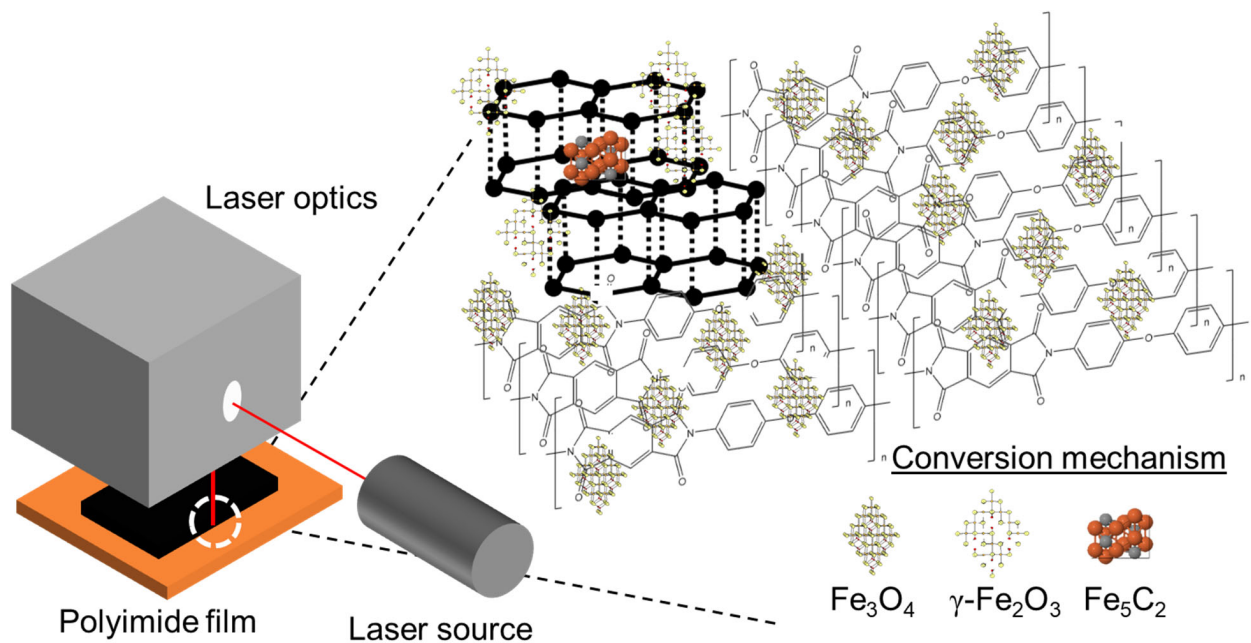

Figure 1. Schematic and conversion mechanism of laser-assisted graphitization on polyimide with magnetite nanoparticles.

**Experimental Methods**

*Synthesis of magnetite nanoparticles:* The magnetite nanoparticles were synthesized by hydrothermal co-precipitation of iron (II, III) salts with NaOH. For the synthesis, 1.75 gm (6.3mmol) of ferrous sulfate heptahydrate ($FeSO_4 \cdot 7H_2O$, Fisher) and 2.92 gm (10.8mmol) of ferric chloride hexahydrate ($FeCl_3 \cdot 6H_2O$, Fisher) powders were added to 50mL of deionized water. The mixture was heated under Argon atmosphere and constant stirring to 80°C using an oil bath. The pH of the mixture was first adjusted to 11 using NaOH, and 0.02g/mL of citric acid was added till pH reached 4.0. The solution was stirred at 80°C for 90 minutes under Ar. Subsequently, the magnetite was precipitated using 1M NaOH, followed by alternating three washing steps with water and acetone. Finally, the collected residue was dried in a vacuum oven at 80°C, at 0.5 atm, and under an Ar blanket for three hours until dry.

*Synthesis of graphitized film and graphene/magnetite nanocomposite:* Kapton polyimide film of 250μm thickness was laser irradiated in the air with 12ps pulsed 10Khz pulse frequency 1064 nm Nd-YAG solid-state laser (Fianium Inc.) to obtain graphitized film. The laser was operated at 2.8W power with a spot diameter of 60μm and a scan speed of 8mm/s.

Graphene/magnetite nanocomposite synthesis was accomplished in three steps. The dried magnetite nanoparticles were crushed into a powder with a mortar pestle and sonicated in methanol for 15 minutes. Three suspensions of magnetite were made with 200 mg of powder in 3 mL (high), 6 mL (medium), and 12 mL (low) methanol. In the first step, 0.5 mL of the high-concentration suspension was applied to the Kapton, and the nanoparticle-covered surfaces were irradiated with a laser power of 0.8W and 8mm/s scan speed. In the second stage, the laser-treated surfaces were coated with 0.5 mL of the medium concentration nanoparticle suspension and irradiated with laser. In the final step, a similar volume of low-concentration nanoparticle suspension was applied to the surface, and the particle-covered surface was again irradiated with a laser at the same processing parameters.

*Materials Characterization:* The magnetite nanoparticles suspended in methanol were characterized for particle size using a Malvern Zetasizer Nano ZS system (DLS). The phase evolution of the magnetite and functionalized nanocomposite was characterized using a Bruker D8 x-ray diffractometer (XRD). The photochemical transformations of the PI to graphene nanocomposite during the laser processing were analyzed using a Kratos Amicus x-ray photoelectron spectrometer (XPS). The degree of graphitization was inspected with a Raman spectrometer (Spectra-Physics **Excelsior** 532-50-CDRH). The morphology and elemental composition of the nanocomposite were imaged with FEI Tenio field emission scanning electron microscope (SEM) with energy dispersive x-ray spectroscopy (EDS). The magnetic property of the nanoparticles and nanocomposite was measured using a Quantum Design SQUID magnetometer at 300K with a maximum applied field of 7T. A four-point probe instrument (Jandel Inc.) was used to characterize the surface conductance of the nanocomposite film.

**Results and Discussions:**

The synthesized MNP had particles with size distributions of 81 ± 40 nm and 138 ± 80 nm and a cumulative distribution of 110 ± 65 nm. The size distribution and the multiple function peak fit (Origin Lab 10) using the Voigt function to delineate two-particle distribution batches of the synthesized nanoparticles are shown in Fig. S1a and Table S1. The MH loop measurements of the synthesized particles, shown in Fig. S1b, indicated a maximum magnetization of 80 emu/g at 7T. The MNP sample had a remanence of 1.56 emu/g and coercivity of 52 Oe, indicating a dominant ferromagnetic behavior and $Fe_3O_4$ phase in the synthesized material.

SEM images of the laser-treated Kapton film are shown in Fig. 2a-c. The picosecond laser used for graphitization has a beam diameter of 60μm. The central ~20μm region of maximum laser intensity ablated the polyimide, creating grooves (Fig. 2a, b) on the top surface. The cross-sectional image of transformed material (including in Fig S2) shows that transformed material has an average thickness of 160 μm and has a closed foam-like porous structure with pore wall thickness of 1μm. This primarily resulted from the rapid electronic excitation and molecular dissociation of the PI chains into layered graphite and in contrast to an open foam porous mesh-like graphene network from $CO_2$ microsecond laser irradiation reported in the literature[7]. The closed foam-like structure indicates that the polymer did not undergo excessive expansion during photolysis, thus maintaining a dense microstructure. A high magnification micrograph of a single layer of the closed foam (Fig. 2c) reveals a folded multi-layered graphene sub-structures with thicknesses of 60 ± 10 nm of individual layers.

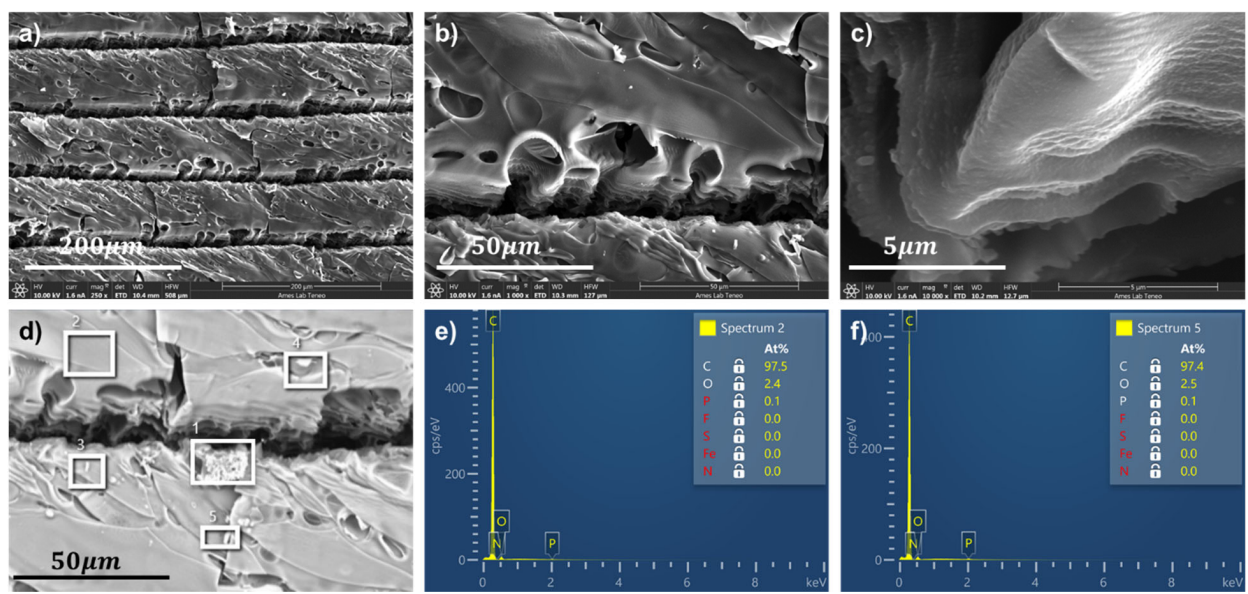

Figure 2. Laser-assisted graphitization of polyimide, a–c) SEM images of treated graphite surface at 250X, 1000X, and 10000X, respectively, d) Cross-section of the L and d – f) EDS point analysis of the treated surface.

The EDS analysis of the treated surface was performed to investigate the elemental composition (Fig. 2d-f and Table S2). The analysis indicated a carbon-dominant structure with no surface nitrogen observed. The small fraction of elemental oxygen was detected and could be attributed to undissociated C-O bonds within PI starting material and atmospheric oxidation during the laser ablation process. However, the relatively small oxygen presence within the sample confirms a successful photochemical dissociation of the PI material. The ultrashort irradiation from the picosecond laser processing minimizes the scope of surface oxidation resulting from the thermal excitation of the lattice. In addition, the nitrogen release from the dissociation of the PI structure may protect the polymer from oxidation during the laser treatment process. An insignificantly small phosphorous content was detected, which could be from sample or SEM chamber contamination.

SEM images of the synthesized MNP-graphite sample are presented in Fig. 3a, b, and the cross-section is shown in Fig S3(a). The SEM images reveal a closed foam porous structure with thinner cell walls compared to laser-treated Kapton film shown in Figure 2. The graphite layers were interlaced with a dispersed network of magnetite nanoparticles and particle agglomerates, as shown in Fig 3(b), (c), and S3b. The cell walls of the composite structure were thinner than the laser-assisted graphite, with a mean thickness of 400 ± 20 nm. A closer inspection of the graphite folds (Fig. 3b) revealed that the MNP particles agglomerate more at the edge than the basal plane. Plausibly, the energy from the laser impact near the center of the beam (ablated area) concentrated the particles over the edge planes. The particle size measured from the SEM images varied from 100 ± 30 nm for the smallest individual spherical nanoparticles to 900 ± 150 nm for the particle agglomerates.

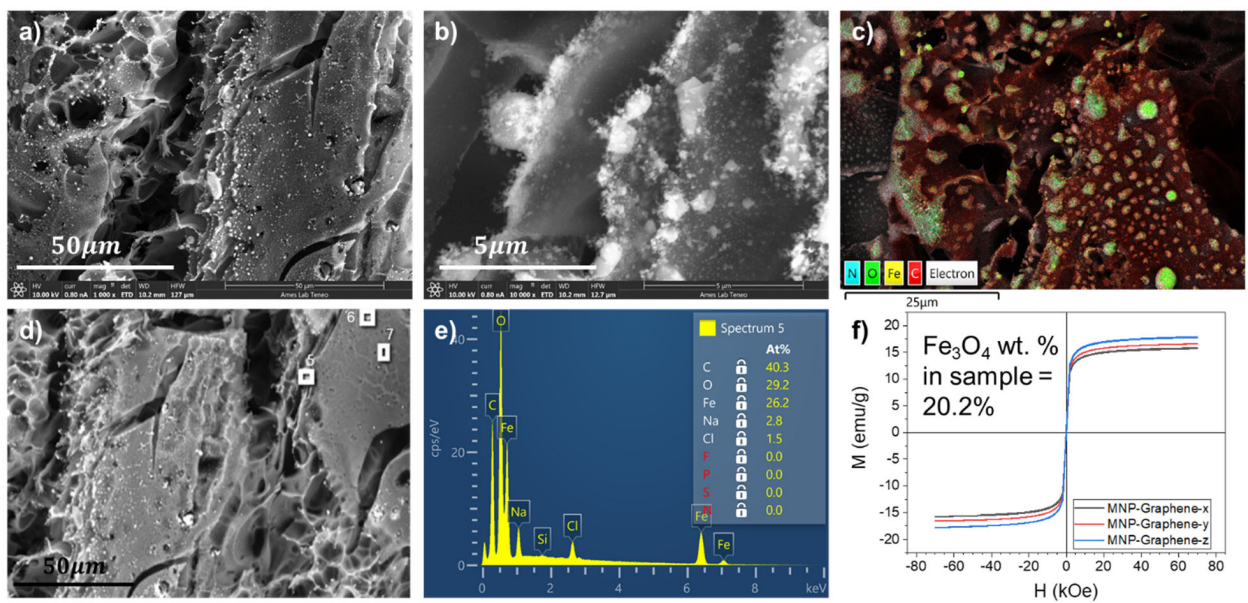

Figure 3. Laser-assisted graphitization of polyimide film functionalized with magnetite nanoparticles, a – b) SEM images of MNP-graphite nanocomposite at 1000X and 10000X, c – e) EDS analysis of the treated surface, EDS map © and point scans (d, e), and f) magnetic characterization (MH) at 300K and 7T.

The EDS map for the cell wall with particle deposition (Fig. 3c) indicates oxygen (primarily) and iron concentration on the particle spots over a carbon background. The nitrogen signal was mainly background with no quantitative indication. The map indicates iron oxide particles embedded on a carbonaceous surface. The strong oxygen signal from the iron oxide particles could be inferred as surface oxidation of the magnetite nanoparticle to a ferric oxide ($Fe_2O_3$) composition. A point-EDS analysis of a lower magnification SEM was performed to observe compositional variations in the composite material (Fig. 3d, e, and Table 1). The primary elemental composition signals were from C, N, O, and Fe, with certain impurities from Na and Cl. In spectrum 5, an almost equal Fe:O ratio was observed, with ~25 at.% Fe and 43 at.% of C. The low O to Fe concentration ratio indicates the elimination of oxygen during laser processing and plausibly iron carbide phase formation. While in spectrum 6 (shown in Table 6), the region primarily contained carbonaceous material (88 at.%) and with a low Fe:O ratio (2:1). The excess O to Fe indicated oxidation of magnetite to ferric oxide with the residual oxygen in the carbon material (similar to Fig. 3c). The absence of N signal in spectrums 5 and 6, indicated a complete dissociation of the PI film. Spectrum 7 (shown in Table 6) primarily contained C, N, and O signals, inferring a possibly unconverted PI impurity in the treated sample. The Na and Cl impurities in spectrum five are acknowledged to be from the MNP synthesis process, where the $Na^+$ (NaOH) neutralizes with $Cl^-$ ($FeCl_3$) to form NaCl.

The magnetic characterization MNP-graphite sample in Fig 3f revealed a ferromagnetic behavior with a maximum magnetization of 17.9 emu/g at 7T, a remnant magnetization of 0.465 emu/g, and a coercivity of 67.8 Oe. Comparing the saturation magnetization value with the as-synthesized MNP (Fig. S1b), the loading fraction from the MNP in the graphite was calculated to be 20.2 wt.%. The small anisotropy observed between the x, y, and z magnetization direction could be due to process-induced anisotropy from laser treatment or a systematic effect from the initial PI film.

Table 1. EDS point scan elemental analysis (wt.%).

| Spectrum Label | C | N | O | F | Na | Cl | Fe | Total |
|---|---|---|---|---|---|---|---|---|
| Spectrum 5 | 43.02 | 0.00 | 28.44 | 0.00 | 2.63 | 1.36 | 24.55 | 100.00 |
| Spectrum 6 | 87.76 | 0.00 | 8.08 | 0.00 | 0.00 | 0.00 | 4.15 | 100.00 |
| Spectrum 7 | 56.69 | 36.64 | 6.71 | 0.26 | 0.00 | 0.00 | 0.00 | 100.00 |

The surface resistance of the electrically conductive film obtained from laser treatment of PI and magnetite/PI nanocomposite films were measured to be $10.0 \pm 0.5$ and $4.0 \pm 0.5$ $\Omega/\square$, respectively. The surface resistances are lower than the reported surface resistance range of 17-28 $\Omega/\square$ for LIG from $CO_2$ lasing [5]. The average thickness of the converted material was computed from cross-sectional images of the two films (shown in Figs. S2 and S3) and resulted in electrical conductivity of $620 \pm 10$ S/m and $1800 \pm 50$ S/m for laser-treated PI and Magnetite/PI films, respectively. Both conductivity measurement and SEM analysis indicate the efficient transformation of PI into an electrically conductive closed foam material by a photochemical dissociation mechanism. Incorporation of the magnetite nanoparticles in the laser-treated material results in a nanocomposite film that is three times higher in electrical conductivity.

The Raman spectrum of the laser-treated PI film is shown in Fig 4(a), labeled "laser-assisted graphite," and reveals graphitization of PI film with an $I_G/I_D$ ratio of 0.79 (<1). The large disorder (D) band indicated incomplete graphitization of the treated sample. A peak between 1700 cm-1 and 2000 cm-1 indicated incomplete conversion with signals for C=O containing impurities in the bulk of the treated sample[19]. Turbostatic disorder and misalignment of graphitic layers could be established from the significant D band at 1340 cm$^{-1}$ and a broad 2D band at 2680 cm$^{-1}$. The Raman spectrum of the laser-treated MNP/PI film is also shown in Fig 4(a) labeled "MNP-graphite, and in comparison, to the laser-assisted graphite, the MNP-graphite sample showed almost six folds improvement in the degree of graphitization, with an $I_G/I_D$ ratio of 4.6. The O impurity and the disorder peaks were almost suppressed. The significant improvements with adding MNP to the PI during laser processing indicated a possible catalytic effect of the magnetite nanoparticles.

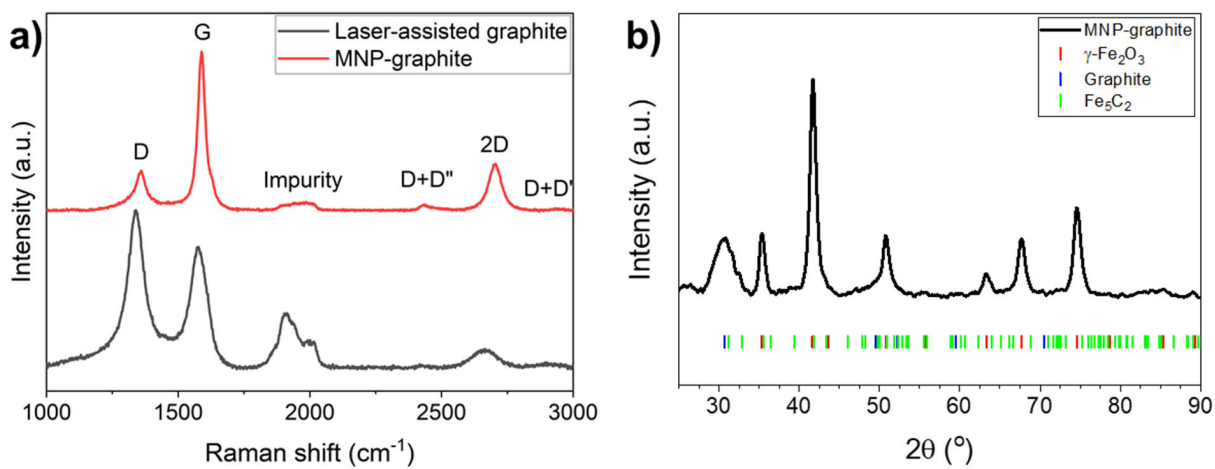

Figure 3. Phase evolution of magnetite during laser photochemical decomposition of polyimide functionalized with MNP, a) Raman spectra of graphite without and with MNP functionalization, and b) XRD of MNP-graphite composite.

XRD characterization of the initial MNP powders (included in supplementary information Fig S5) and post-laser treatment MNP-graphite film shown in Fig 4b were analyzed to understand the phase evolution during the laser treatment. The initial MNP as-synthesized powder (XRD results shown in Fig. S3) is composed of magnetite ($Fe_3O_4$, Pearson 1816552[20]) and maghemite ($\gamma$-$Fe_2O_3$ Pearson 452633[21]) phases. Analysis of the peaks revealed the sample was composed of 95 wt.% of magnetite and five wt.% maghemite (Table 2). The compositional analysis supported the magnetic characterization of the powders (shown in Fig S1(b)), wherein the high magnetization was predominantly from the magnetite phase in the MNP powder. XRD spectrum of the magnetite-graphite composite revealed peaks from several phase compositions, including graphite, maghemite, and iron (II, III) carbide. The primary graphite peaks were at 30.72º and 52.14º (Pearson 1817309[22]). Although magnetite and maghemite peaks are relatively close in XRD, the observed lattice parameter, by refining the fitted curve for the cubic crystal, was estimated to be 8.34Å (maghemite) compared to magnetite (at 8.39Å). The primary maghemite peaks fitted with the spectrum were 35.28º, 41.64º, 50.76º, 67.66º, and 74.62º. In addition, several $Fe_5C_2$ (Pearson 1617122[23]) peaks located at 39.42º, 41.82º, 43.32º, 46.06º, 47.80º, 48.20º, 50.08º, 50.94º, 51.78º, 52.38º, 52.84º, 55.5º, 59.06º, 60.64º, and 68.86º were observed. Refinement and analysis of the MNP-graphite spectrum suggested 21.5 wt.% of maghemite ($\gamma$-$Fe_2O_3$), 4.0 wt.% of iron carbide ($Fe_5C_2$), and 74.5 wt.% of graphite with residual polyimide (Table 2). The calculated composition from the XRD spectrum matches within 6% of the estimated magnetic loading fraction from the MH characterization (Fig. 3f).

Table 2. Phase and compositional analysis of XRD for magnetite nanoparticles and MNP-graphite composite.

| Phase | MNP (wt.%) | MNP-Graphite (wt.%) |
|---|---|---|
| $\gamma$-$Fe_2O_3$ | 5.7 | 21.5 |
| $Fe_3O_4$ | 94.3 | 0.0 |
| $Fe_5C_2$ | 0.0 | 4.0 |
| Graphite + PI | 0.0 | 74.5 |

XPS analysis of MNP-graphite surface composite revealed elemental compositions containing primarily C, O, N, and Fe (Spectrum and fits shown in Fig. S4 and analysis results in Table S3). Deconvolution of the wide spectrum revealed a carbonaceous surface (37.7 at.%) with iron (18.55 at.%) and oxygen (42.28 at.%) species. The atomic ratio Fe to O suggests Fe (III) or maghemite composition with excess surface oxygen from impurities in the decomposed carbon material.

Narrow spectra of the MNP-graphite film surface corresponding to C, O, Fe, and N elements and their corresponding deconvolutions are shown in Figs 5a,b,c, and d, respectively.

The C1s spectrum (Fig. 5a) was fitted with sp$^2$ ($C = C$) at 284.7 eV, and respective $\pi - \pi^*$ satellite peak at 288.91 eV, $C - O$ at 286.45 eV, $C = O$ at 288.5 eV and carbide at 283.1 eV[24–27]. The C 1s spectrum analysis indicates the composite surface was composed primarily of graphite (77.5 at.%), with 12.5 at.% of carbon-oxygen species and 9.9 at.% of carbide of the C 1s composition. From the O 1s (Fig. 5b), 38.8 at.% of O 1s was $O^{2-}$ from $C = O$ & γ-Fe$_2$O$_3$ (530.3 eV), and 62.2 at.% of O 1s was O$^{1-}$ or $C - O$ (at 531.2 eV)[28,29]. In the Fe 2p spectrum (Fig. 5c), peaks were located at 711.0 eV and 724.5 eV, equivalent to γ-Fe$_2$O$_3$ (2p-3/2 and 2p-1/2) and their respective $\pi - satellite\pi^*$ satellite peaks at 714.0 eV and 728.1 eV, respectively[30]. In addition to γ-Fe$_2$O$_3$, peaks for iron carbide (Fe$_5$C$_2$) were also located at 706 eV and 719.2 eV[31]. Deconvolution of the peaks resulted in γ-Fe$_2$O$_3$ of 91.2 at.% of Fe 2p, and remaining as iron carbide (Fe$_5$C$_2$) species[32]. Therefore, the XPS analysis suggested that the treated MNP-graphite composite surface was composed of a graphitic structure with maghemite and iron carbide as the magnetic phases and carbon-oxygen species as the unconverted impurities.

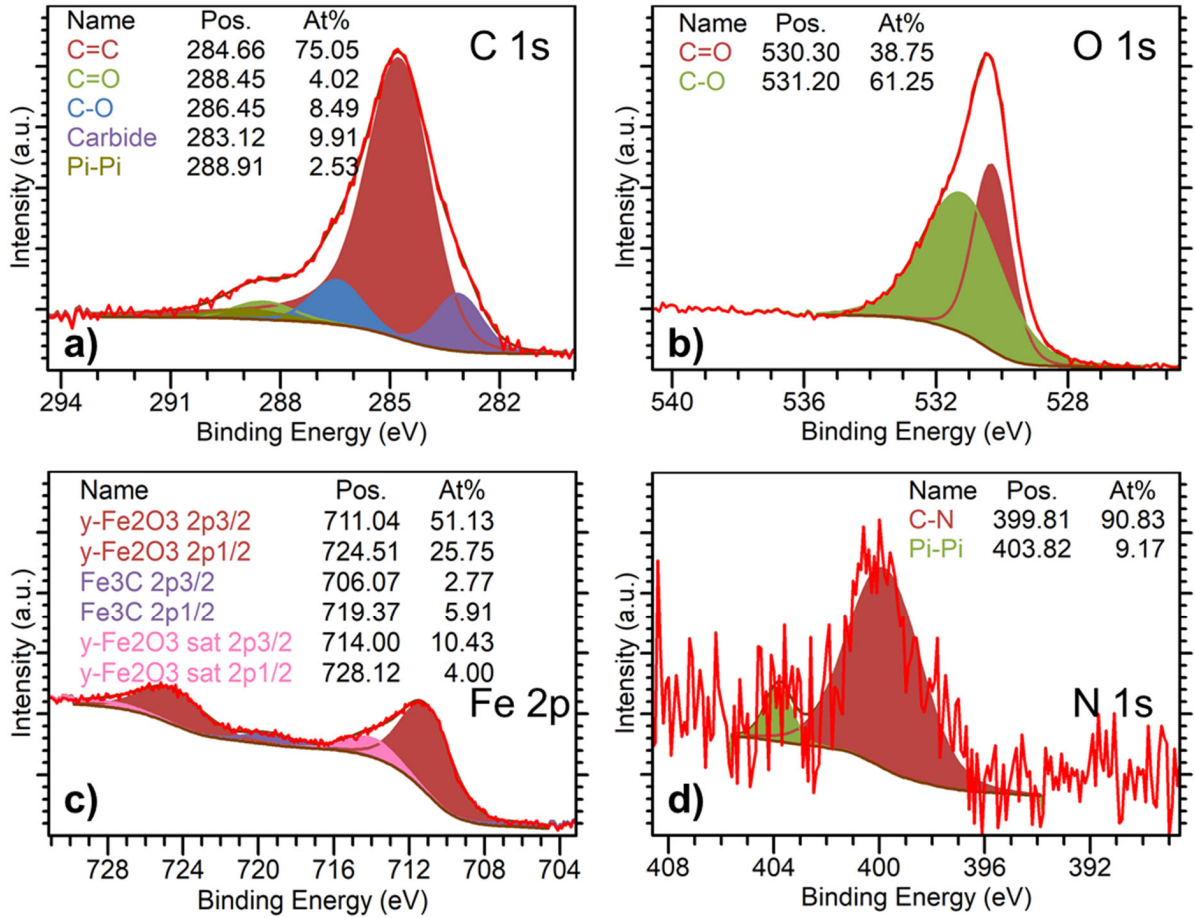

Figure 4. XPS narrow spectra and chemical compositional analysis of treated MNP-graphite composite, a) C 1s, b) O 1s, c) Fe 2p, and d) N 1s.

The morphological and chemical characterization of the laser-treated PI and MNP-graphite film surface indicates that the laser graphitization of polyimides under ultrashort pulsed laser irradiation follows a photolytic transformation mechanism. Initially, the incident laser pulse interacts and couples to an electron system, transferring the electromagnetic energy as kinetic (thermal) and potential (band gap) energy. Typically, the electron-lattice relaxation time is in the order of picoseconds ($10^{-12}$ s)[11]. In picosecond lasers, the laser pulse width of the beam transferred to the electrons is shorter than the electron-lattice relaxation period. The potential energy in the band gap does not get the opportunity to convert into kinetic energy (via the Auger process) and dump into the lattice via electron-ion collisions to bring the system to a thermal equilibrium.

Instead, the energy remains within the electron gas and is not dissipated within the lattice. Since the electron gas relaxation time is $10^{-14}$ s[18] (< laser pulse width), the electron system (at >10,000K) equilibrates itself but remains at a significantly higher temperature than the lattice (at 300K)[6]. The polyimide chain, consisting of benzene rings with associated imide functional groups (Fig.1a), acts as an excellent starting material due to the ring structure and presence of nitrogen-containing molecules. The ring structure from the benzene (hexagon) and imide (pentagon) re-organize into $sp^2$ hybridized graphitic structure during photodecomposition. The N within the imide group dissociates into free radicals within the plasma, either absorbing the evolving oxygen radical to form $NO_x$ or forming molecular $N_2$ and protecting the treated region from further oxidation. The higher electronic excitation of picosecond pulsed lasers allows for rapid molecular dissociation without distorting the lattice structure and enables a layered graphitic material that forms the walls of the closed foam structure observed in the SEM micrographs.

The chemical phase evolution characterized using XPS, EDS, and XRD suggests that magnetite nanoparticles convert to maghemite and iron carbide, acting as nucleation sites that couple laser irradiation onto the polyimide surface and catalyzing the photolysis mechanism. The MNP accelerates the oxygen elimination of the C structure during conversion. Naturally, the maghemite is formed due to vacancies created by ejection of $Fe^{2+}$ from the magnetite crystal lattice and represented as $(Fe_8^{III})_A \left[ Fe_{\frac{40}{3}}^{III} \square_{\frac{8}{3}}^{II} \right]_B O_{32}$, where the $Fe^{II}$ atoms are vacant in the octahedral sites. During the laser conversion process, the Fe from magnetite reacts with the carbon to form the iron carbide phase ($Fe_5C_2$), thus reducing the magnetite to maghemite. This process binds the carbon to the nanoparticles while eliminating the N and O species into the plasma plume. The laser irradiation of PI surface in the presence of MNP results in a higher degree of graphitization and higher electronically conductive films than laser irradiation of PI alone. These results suggest that nanoparticles play a significant role as a photocatalyst to accelerate the photolytic decomposition process and get incorporated into the graphitized materials forming magnetite/graphite nanocomposite.

**Conclusion**

We report the photochemical decomposition of magnetite-functionalized graphite nanocomposite from the photochemical decomposition of polyimide film in the presence of magnetite nanoparticles using a picosecond pulsed laser. Utilizing ultrashort pulsed lasers induces photochemical transitions in polymers, resulting in a graphitic film with excellent electronic conductivity and the morphology of closed foam enclosed within planar surfaces. Irradiation of magnetite nanoparticle-covered PI surface resulted in an MNP-graphite nanocomposite with higher electronic conductivity and closed foam morphology where magnetite nanoparticles and particle agglomerates are dispersed in the cell walls. Magnetic measurements of the nanocomposite indicated ferromagnetic behavior with process-induced magnetic anisotropy. Raman spectra of the MNP-graphite composite had a four times higher degree of graphitization than laser-converted graphite. Additionally, A fourfold increase in electrical conductivity was measured due to magnetite nanoparticle inclusion in the MNP/graphite nanocomposite. Characterization results suggest that magnetite nanoparticles suppress the N impurities and turbostatic disorders of the graphitic structures formed during laser irradiation, suggesting a possible photocatalytic mechanism. EDS, XRD, and XPS analysis suggested that the magnetite nanoparticles acted as nucleation sites to distribute the laser energy for rapid photochemical conversion of PI film. The magnetite bonded with the carbon chain by forming iron carbide ($Fe_5C_2$) and eliminated O from the polymer structure by reducing to maghemite ($\gamma$-$Fe_2O_3$). Therefore, the magnetite nanoparticles are not only incorporated into the graphite to form a magnetic nanocomposite with higher economic potential but also catalyze the photochemical dissociation process for an energy-efficient laser graphitization.

## Data Availability

The raw/processed data required to reproduce these findings cannot be shared at this time as the data also forms part of an ongoing study but will be made available on request to the corresponding author.

## Acknowledgement

This work was supported by the Department of Energy, Laboratory Directed Research and Development program at Ames Laboratory. Ames Laboratory is operated for the U.S. Department of Energy by Iowa State University of Science and Technology under Contract No. DE-AC02-07CH11358.